\documentclass[12pt,a4paper]{article} 
\usepackage{amsmath}
\usepackage{amssymb} 
\usepackage{graphicx} 
\usepackage{float}
\restylefloat{table}
\begin{document}
\begin{titlepage} 
\begin{flushright} IFUP--TH/2013-14\\ 
\end{flushright} ~
\vskip .8truecm 
\begin{center} 
\Large\bf Hyperbolic deformation of the strip-equation and the
accessory parameters for the torus
\end{center}
\vskip 1.2truecm 
\begin{center}
{Pietro Menotti} \\ 
{\small\it Dipartimento di Fisica, Universit{\`a} di Pisa and}\\ 
{\small\it INFN, Sezione di Pisa, Largo B. Pontecorvo 3, I-56127}\\
{\small\it e-mail: menotti@df.unipi.it}\\ 
\end{center} 
 \vskip 1.2truecm
\centerline{July 2013}
                
\vskip 1.2truecm
                                                              
\begin{abstract}

By applying an hyperbolic deformation to the uniformization problem
for the infinite strip, we give a method for computing the accessory
parameter for the torus with one source as an expansion in the modular 
parameter $q$. At $O(q^0)$ we obtain the same equation for the accessory
parameter and the same value of the semiclassical action as the one
obtained from the $b\rightarrow 0$ limit of the quantum one point
function. The procedure can be carried over to the full $O(q^2)$ or 
even higher order corrections
although the procedure becomes somewhat complicated.
Here we compute to order $q^2$ the correction to the weight parameter 
intervening in the conformal factor and it is shown that the unwanted 
contribution $O(q)$ to the accessory parameter equation cancel exactly. 
\end{abstract}

\end{titlepage}

\eject

\section{Introduction}

~~~The accessory parameters related to punctured Riemann surfaces
play an important role in conformal field theories. Not only they
give an explicit solution to the uniformization problem but through
the Polyakov relation they provide the dependence of the action,
i.e. of the semiclassical limit of the quantum correlation functions,
on the position of the singularities. A lot of work has been devoted
to the determination of such accessory parameters which turned out to
be a highly transcendental problem.

Several conjectures and proposals
\cite{ZZ,hadaszjaskolski1,hadaszjaskolski2,hadaszmodular,ferraripiatek}, 
\cite{FLNO},\cite{torusI,torusII,torusIII},\cite{poghossian},
\cite{troost} for the 
computation of such accessory parameters have been put forward
e.g. taking the semiclassical limit of the
quantum correlation functions in conformal theories, 
the 4-point function on the sphere and
the 1-point function on the torus . More recently
a renewed interest in conformal blocks, which are related to the
accessory parameters, arose in connection with the AGT conjecture
\cite{gaiotto,AGT} that Liouville theory on a Riemann
surface of genus $g$ is related to a certain class of $N = 2$,
four-dimensional gauge theories and the conjecture has been supported
by extensive tests on genera 0 and 1 \cite{AGT,albamorozov,drukker} 
and proven in a class of cases \cite{hadaszmodular,hadaszproving}.

A better understanding of the classical counterpart would shed light
on several problems concerning the conformal blocks e.g. the
convergence region of the expansion in the invariant cross ratio, 
or in the ``nome'' $q=e^{i\pi\tau}$ for the torus and the validity of the
exponentiation hypothesis \cite{ZZ}.

Most of the literature, in particular the mathematical literature
\cite{KRV,kra,smithhempel,hempelsmith,hejhal,hempel2,hempel3} 
is concerned with
parabolic singularities i.e. punctures. The reason lies in the fact
that in this case results from the fuchsian mapping theory can be
applied. On the other hand in Liouville theory the case of
elliptic singularities is of most interest.

From the general viewpoint it was proven in \cite{torusIII} that the
accessory parameter for the torus with one elliptic singularity 
is an analytic function of the coupling in the
whole physical range except at most a finite number of points and that
the accessory parameter is a real-analytic (not analytic) function  of
the complex modulus $\tau$ except at most a zero measure set in the
fundamental region for $\tau$, extending results of 
\cite{CMSphyslett,CMS}.

In addition it was found that the perturbative series in the source
strength converges in a finite region and a rigorous lower bound on
the convergence radius given \cite{torusIII}.
Moreover the first \cite{torusI} and second order expression
\cite{torusIII} for the accessory parameter was explicitly written in
terms of elliptic functions and related integrals.

In conformal theories an other type of expansion has been brought to
attention i.e. the expansion in the modular parameter which appears in
the formal expression of conformal blocks.

For the four point problem on the sphere the expansion of the
conformal blocks is done in the invariant cross ratio $x$ which
corresponds to an expansion in the distance of a pair of
singularities after the others have been fixed in standard way.
The conformal blocks are defined as a formal power expansion in such
parameter but little is known about the convergence of such a series
even though numerical investigations give good support for a
convergence region common to the parameters $x$ and $1-x$ where the
validity of the conformal bootstrap can be numerically verified
\cite{ZZ}, \cite{hadaszrecursive}.

Similar problems occur for the torus with one source.
Important results regarding the four-point
conformal correlation functions and their relation to the one-point
function on the torus have been obtained in \cite{FLNO} and
\cite{hadaszmodular}.

The dependence of the accessory parameter from the distance, in the
limit of coalescing singularities has been studied rigorously for
parabolic singularities in simplified mathematical models
\cite{smithhempel,hempelsmith}. At the semiclassical level it was
suggested that one can reach the accessory parameters through the
ansatz of the exponentiation of the conformal blocks in the classical
limit \cite{ZZ,ferraripiatek} and a saddle point procedure in such a
limit.

The present paper examines the dependence of the accessory parameter
for the torus with a single source
for small values of $|q|$ directly from the classical Liouville theory.
The final aim is to obtain a direct comparison with the
procedures described above, which start from the quantum correlation
function.

In papers \cite{torusI,torusII,torusIII} it was proven that in addition
to the case of the square and the equianharmonic case i.e. the torus
with $\tau = e^{i\pi/3}$, the limit case of the infinite
strip is soluble. This 
correspond to $q=e^{i\pi\tau}=0$. The idea developed in the present
paper is to treat
large but finite values of ${\rm Im}\tau$ starting form the soluble
infinite strip problem through an appropriate deformation or the
relative equations.  This is reached by an expansion in $q^2$ 
of the kernel given by the Weierstrass $\wp(z)$ function whose first
two terms in fact represent the problem for the strip. The procedure is to
compute the three fundamental monodromies and to impose on them the
$SU(1,1)$ nature which is the necessary and sufficient condition for
the single valuedness of the conformal factor. The developed treatment
although perturbative in $q$ is completely non perturbative in the
source strength $\eta$ and thus can be applied to the whole
physical range $0<\eta\leq 1/2$.

The paper is structured as follows. In section \ref{general}
we give the general description of the problem and
summarize some results on the treatment of the strip given in
\cite{torusIII} which will be necessary for the subsequent developments.
In section \ref{deformation} we examine all possible deformations of
the equation for the strip proving that  even
though elliptic deformations can satisfy the requirement for the 
monodromy around the source
and along the short cycle, they cannot satisfy the $SU(1,1)$ requirements 
on the long cycle. The hyperbolic deformation on the other hand is 
consistent and
provides, working to order $O(q^0)$, an implicit equation for the
accessory parameter. Integrating an equation in $\eta$ one obtains
also the value of the action i.e. of the semiclassical $1$-point
function. We show in the same section that the derived equation
for the accessory parameter coincides with the one obtained from the 
saddle point treatment of the quantum $1$-point function and that the
derived action equals the quantum action in the semiclassical limit.

In section \ref{discussion} we perform the first iteration of the
Volterra integral equation in which the $q^2$ term in the expansion of
the kernel is taken into account. The change due to this term of the
weight parameter appearing in the conformal factor which is necessary
to impose the consistency of the monodromies, is computed.

On the other hand due to the exponential behavior of the kernel for
large imaginary values of the coordinate it is shown that the first 
term in the expansion of the kernel which is formally $O(q^2)$ 
contributes actually $O(q)$ to the computation of the monodromy on the 
long cycle. It is shown that this does not contrast with the structure of
the quantum correlation function, as such contribution cancel exactly
with the term of the same order appearing in the expansion of the
unperturbed function.  However due to this fact to reach the full
$O(q^2)$ order correction to the equation for the accessory parameter
one has to perform two iterations of the integral equation taking into
account also the second term in the expansion of the kernel. This is a
rather lengthy process which will be pursued in an other work. In
general to reach order $q^{2n}$ one has to perform not $n$ but $2n$
iterations in the integral equation.

\section{General setting and the infinite strip problem}\label{general}

Liouville equation with one generic source for the torus 
\cite{KRV} 
\begin{equation}\label{liouvilleeq}
-\partial_z\partial_{\bar z}\phi+e^{\phi}= 2\pi \eta~\delta^2(z)
\end{equation}
can be translated into the ordinary differential equation
in the complex plane
\begin{equation}\label{zdiffeq}
{f^t}''(z)+\epsilon (\wp(z)+\beta)f^t(z)=0
\end{equation}
with
\begin{equation}\label{eminusphiovertwo}
e^{-\phi/2} =\frac{1}{\sqrt{2}|w_{12}|}
\big[\kappa^{-2}\overline{f^t_1(z)} f^t_1(z)-
\kappa^2 \overline{f^t_2(z)} f^t_2(z)\big]
\end{equation}
being $f^t_1(z)$ and $f^t_2(z)$
two independent solutions of eq.(\ref{zdiffeq}), $w_{12}=
f^t_1(z){f_2^t}'(z)-{f_1^t}'(z)f^t_2(z)$. $\kappa$
is a real weight parameter and $\beta$ is the accessory parameter to 
be determined as to satisfy the torus periodic boundary conditions.
The approach pursued in the present paper is to exploit the convergent
expansion for $\wp(z)$
\cite{DLMF}
\begin{equation}\label{wpexpansion}
\wp(z) =-\frac{\zeta(\omega_1)}{\omega_1}+\frac{\pi^2}{4\omega_1^2}
\frac{1}{\sin^2\big(\frac{\pi z}{2\omega_1}\big)}-\frac{2\pi^2}{\omega_1^2}
\sum_{n=1}^{\infty}\frac{nq^{2n}}{1-q^{2n}}\cos\big(\frac{n\pi z}{\omega_1}\big)
\end{equation}
where $q=e^{i\pi\tau}$ is the ``nome'' of the torus, $\tau$ its modulus
and $\omega_1$ the first half-period. 

The interest for such an expansion is that the problem where only the
first two terms in eq.(\ref{wpexpansion}) are taken into account is
exactly soluble \cite{torusIII} and thus the
remaining terms can be treated as a perturbation. The unperturbed
problem is the infinite strip problem.

In \cite{torusIII} the conformal factor for the infinite strip with one
source was explicitly given in terms of hypergeometric functions.
Due to the relevance of this result for the following treatment, we
report here the main formulas and notations.

The infinite vertical strip is a degenerate case of the tours 
\cite{batemanII}
reached with the parameters
$e_1=2a$, $e_2=e_3 =-a$. The Weierstrass
$\wp$ and $\zeta$ functions associated with the torus degenerate to
\begin{equation}\label{degeneratewpandzeta}
u=\wp(z)=-a+\frac{3}{\sin^2(\sqrt{3a}~z)},~~~~~~~~
\zeta(z) = az +\sqrt{3a}~\frac{\cos(\sqrt{3a}~z)}{\sin(\sqrt{3a}~z)}
\end{equation}
with
\begin{equation}
\omega_1 =\frac{\pi}{2\sqrt{3a}},~~~~~~~~\omega_2 = i\infty~.
\end{equation}
In the following we shall normalize $a=1$. 
The differential equation associated with the uniformization problem in
presence of a source at the origin $z=0$ is
\begin{equation}\label{udiffeq}
y''(u)+ Q(u) y(u)=0
\end{equation}
with
\begin{equation}\label{stripQ}
Q(u) = \frac{1-\lambda^2}{16}\frac{u+\beta}{(u+1)^2(u-2)}+
\frac{3}{16}\frac{u^2-2u+9}{(u-2)^2(u+1)^2}~.
\end{equation}
and  
\begin{equation}
\eta =\frac{1-\lambda}{2},
~~~~\epsilon = \frac{1-\lambda^2}{4}~.
\end{equation}
The limit of an infinite rectangle requires $\beta=1$ \cite{torusIII}. 
The first and second order values of $\beta$ as a power expansion in
$\eta$ for the general torus
were given in \cite{torusI,torusII,torusIII}. To first order
we have
\begin{equation}
\beta =\frac{\zeta(\omega_2)\bar \omega_1-
\zeta(\omega_1)\bar \omega_2}{\omega_2\bar\omega_1-\omega_1\bar\omega_2}~.
\end{equation}
It is of interest that  using (\ref{degeneratewpandzeta}) one gets
$\beta=1$ already to first order
\begin{equation}\label{betastrip}
\lim_{\omega_2\rightarrow i \infty}~\beta =1~.
\end{equation}

Two independent solutions
of the differential equation (\ref{udiffeq})  are 
\begin{equation}
y_1 =(-v)^{1/4}(1-v)^{1/2} ~{_2F_1}(\frac{1-\lambda}{4},\frac{1+\lambda}{4};
\frac{1}{2};v)
\end{equation}
\begin{equation}
y_2 =2 (-v)^{3/4}(1-v)^{1/2} ~{_2F_1}(\frac{3-\lambda}{4},\frac{3+\lambda}{4};
\frac{3}{2};v)
\end{equation}
with 
\begin{equation}
v = \frac{2-u}{3}~.
\end{equation}
Such a pair of solutions is canonical at $v=0$, thus
assuring monodromy at the point $v=0$ which corresponds to
$z=\omega_1$.
The functions $f_1(z)$ and $f_2(z)$ relative to the original problem
are obtained through the standard transformation, taking into account
their nature of order $-1/2$--forms \cite{hawley}, using
\begin{equation}\label{dvdz1}
\big(\frac{dv}{dz}\big)^{-\frac{1}{2}} = \frac{e^{\frac{i\pi}{4}}}
{\sqrt{2} ~3^{1/4}(-v)^\frac{1}{4}
(1-v)^\frac{1}{2}}
\end{equation}
while $\kappa$ is obtained by imposing the $SU(1,1)$ nature of the
monodromy at $u=\infty$ which corresponds to $z=0$ to obtain \cite{torusIII}
\begin{equation}
\kappa^4=\bigg[\gamma(\frac{3-\lambda}{4})
\gamma(\frac{3+\lambda}{4})\bigg]^2
\end{equation}
where as usual
\begin{equation}
\gamma(x)=\frac{\Gamma(x)}{\Gamma(1-x)}.
\end{equation}
The quantity $X_z$ appearing in the expansion
\begin{equation}
\phi(z) = -2\eta\log|z|^2+X_z+o(z)
\end{equation}
is given by
\begin{equation}
X_z = \log 6+ 2 \eta \log\frac{4}{3} -2\log\gamma(\eta)-
2\log \gamma(\frac{1}{2}-\eta)~.
\end{equation}
Integrating the equation
\begin{equation}
\frac{1}{2\pi}\frac{\partial S_z({\rm strip})}{\partial \eta} = -X_z
\end{equation}
we have for the action
\begin{equation}
\frac{1}{2\pi}S_z({\rm strip})= 
-\eta \log 6 -\eta^2\log\frac{4}{3}+2 F(\eta)-2 F(\frac{1}{2}-\eta)-2 F(0)
\end{equation}
where
\begin{equation}\label{F}
F(x) = \int_{\frac{1}{2}}^x \log \gamma(x') dx'.
\end{equation}

\bigskip
\section{\bf Deformation of the equation for the strip}\label{deformation}
We shall denote by $C_1$ the cycle encircling the origin
$z=0$ which is the location of the source, with $C_2$ the cycle 
obtained by identifying $z$ with $z-2\omega_1$ (the short cycle)
while the cycle $C_3$ (the long cycle) is the one obtained by
identifying the line ${\rm Im}~z=\omega_1\tau_2$ with the line 
${\rm Im}~z=-\omega_1\tau_2$. The procedure we shall employ to 
determine the accessory parameter for
the elongated rectangle is to impose in addition to the monodromy
conditions on the cycle $C_1$  and $C_2$ the monodromy on the cycle
$C_3$ i.e. the periodic conditions at ${\rm Im}(z) = \pm \omega_1\tau_2$.

This, as we shall see, is consistent only with values of the
accessory parameter $\beta\neq 1$, and here we examine the nature of
such a deformation.

Separating the soluble part of eq.(\ref{zdiffeq}) from the remainder 
we can translate the original differential equation to
the Volterra type integral equation
\begin{equation}\label{volterra}
f^t_k(z) = f_k(z) +\epsilon \int_{\omega_1}^z K(z')G(z,z')f^t_k(z')dz',~~~~k=1,2
\end{equation}
where
\begin{equation}
G(z,z') = \frac{1}{w_{12}}(f_1(z)f_2(z')-f_2(z)f_1(z'))
\end{equation}
and
\begin{equation}\label{Kexpansion}
K(z') =-\frac{2\pi^2}{\omega_1^2}
\sum_{n=1}^{\infty}\frac{nq^{2n}}{1-q^{2n}}\cos\big(\frac{n\pi z'}{\omega_1}\big)
\equiv \sum_{n=1}^\infty K_{n}(z')
\end{equation}
and $f_k$ are the solutions for $q=0$.
Eq.(\ref{volterra}) can be solved by the standard convergent 
iteration procedure.
Notice that the chosen lower integration bound assures that the
solutions $f^t_k(z)$ of eq.(\ref{volterra}) are still canonical at 
$z=\omega_1$ thus
assuring to all orders the single valued behavior of the conformal
factor around $\omega_1$.

It is of interest that the terms $\cos(n\pi z/\omega_1)$ in the 
expansion (\ref{Kexpansion})
formally contribute starting from order $O(q^{2n})$. It will not so in
practice due to exponential behavior of the cosine for large imaginary
values of $z$ thus contributing $O(q^n)$ to the evaluation of
the monodromy $C_3$. This is uncomfortable as to reach the precision
$O(q^{2n})$ we have to take into account $2n$ terms in the expansion
(\ref{Kexpansion}) and accordingly $2n$ iterations of
(\ref{volterra}). This will be discussed in detail in section 
\ref{discussion}.

On the other hand in the evaluation of the monodromies $C_1$ and $C_2$ the
$n$--th term contributes starting from the order $q^{2n}$ as one would
naively expect.
 
 We saw that the problem for the strip is
solved by $\beta=1$. For $\beta = 1 +\beta_1$,
$\beta_1\neq 0$ a pair of solution $f_k(z)$
is given by
\begin{equation}
f_1(z) =(1-v)^{\Lambda}~
_2F_1(\frac{1-\lambda}{4}+
\Lambda,\frac{1+\lambda}{4}+
\Lambda;\frac{1}{2};v)
\end{equation}
\begin{equation}
f_2(z) = 2~(-v)^\frac{1}{2}(1-v)^{-\Lambda}~
_2F_1(\frac{3-\lambda}{4}-\Lambda,\frac{3+\lambda}{4}-
\Lambda;\frac{3}{2};v)
\end{equation}
where
\begin{equation}
\Lambda = \frac{1}{4}\sqrt\frac{(1-\lambda^2)\beta_1}{3}~.
\end{equation}
In the rest of this section we shall work to level $O(q^0)$.

In order to compute the monodromy along $C_1$ we need the behavior of
$f_1,f_2$ at $v=\infty$. This is given by
\begin{equation}
f_1 = B^{(1)}_1 (-v)^{-\frac{1-\lambda}{4}} +
B^{(1)}_2 (-v)^{-\frac{1+\lambda}{4}}
\end{equation}
\begin{equation}
f_2 = B^{(2)}_1 (-v)^{-\frac{1-\lambda}{4}} +
B^{(2)}_2 (-v)^{-\frac{1+\lambda}{4}}
\end{equation}
with
\begin{equation}
B^{(1)}_1 = \frac{\Gamma(\frac{1}{2})\Gamma(\frac{\lambda}{2})}
{\Gamma(\frac{1+\lambda}{4}-\Lambda)\Gamma(\frac{1+\lambda}{4}+\Lambda)},
~~~~~~~~
B^{(1)}_2 = \frac{\Gamma(\frac{1}{2})\Gamma(-\frac{\lambda}{2})}
{\Gamma(\frac{1-\lambda}{4}-\Lambda)\Gamma(\frac{1-\lambda}{4}+\Lambda)}
\end{equation}
\begin{equation}
B^{(2)}_1 = \frac{\Gamma(\frac{3}{2})\Gamma(\frac{\lambda}{2})}
{\Gamma(\frac{3+\lambda}{4}-\Lambda)\Gamma(\frac{3+\lambda}{4}+\Lambda)},
~~~~~~~~
B^{(2)}_2 = \frac{\Gamma(\frac{3}{2})\Gamma(-\frac{\lambda}{2})}
{\Gamma(\frac{3-\lambda}{4}-\Lambda)\Gamma(\frac{3-\lambda}{4}+\Lambda)}~.
\end{equation}
The monodromy matrix for a complete turn in $z$ is
\begin{equation}\label{M(C1)}
M(C_1)=-\lambda
\begin{pmatrix}
(B^{(1)}_1B^{(2)}_2e^{i\pi(1-\lambda)}-B^{(1)}_2B^{(2)}_1e^{i\pi(1+\lambda)})&
B^{(1)}_1B^{(1)}_2(e^{i\pi(1+\lambda)}-e^{i\pi(1-\lambda)})/2\\
2~B^{(2)}_1B^{(2)}_2(e^{i\pi(1-\lambda)}-e^{i\pi(1+\lambda)})&
(B^{(1)}_1B^{(2)}_2 e^{i\pi(1+\lambda)}-B^{(1)}_2B^{(2)}_1 e^{i\pi(1-\lambda)})
\end{pmatrix}
\end{equation}
from which the value of the parameter $\kappa^4$ appearing in
eq.(\ref{eminusphiovertwo}) is derived
using $\kappa^4=M_{12}(C_1)/\overline{M_{21}(C_1)}$
\begin{equation}\label{kappa4}
\kappa^4 =
\frac{\overline{\Gamma(\frac{3-\lambda}{4}-\Lambda)
\Gamma(\frac{3-\lambda}{4}+\Lambda)
\Gamma(\frac{3+\lambda}{4}-\Lambda)\Gamma(\frac{3+\lambda}{4}+\Lambda)}}
{\Gamma(\frac{1-\lambda}{4}-\Lambda)
\Gamma(\frac{1-\lambda}{4}+\Lambda)
\Gamma(\frac{1+\lambda}{4}-\Lambda)\Gamma(\frac{1+\lambda}{4}+\Lambda)} ~. 
\end{equation}
We recall that in order to have a single valued conformal factor such
$\kappa^4$  should be real and positive. From the previous
expression we see that reality is achieved only for
$\Lambda$ real i.e. elliptic deformation, or $\Lambda$ pure imaginary
i.e. hyperbolic deformation. For $\Lambda$ real, $\kappa^4$ is
positive only for small values of $|\Lambda|$. On the other
hand for $\Lambda$ pure imaginary $\kappa^4$ is always real and
positive. Moreover due to the reality of the
$B^{(j)}_k$ for real or imaginary $\Lambda$, we see from eq.(\ref{M(C1)})
that $M_{22}(C_1)=\overline{M_{11}(C_1)}$ thus assuring that the
matrix $M(C_1)$ transformed by ${\rm diag}(\kappa^{-1},\kappa)$ belongs to
$SU(1,1)$. Thus we have single valuedness of the
conformal factor around the source for any imaginary $\Lambda$ and not too
large real $\Lambda$. We shall see later that real values of $\Lambda$
are excluded by an other reason. For a rotation of $\pi$ in $z$ and from the 
invariance of eq.(\ref{zdiffeq}) we have a $U(1,1)$ monodromy matrix 
which assures that the conformal factor is invariant under 
such a transformation.

We come now to the imposition of
the monodromy condition along the cycle $C_2$, obtained by identifying
${\rm Re}~z=\omega_1$ with ${\rm Re}~z=-\omega_1$.
To deal with this problem it is useful to rewrite the
functions $f_1(z)$ and $f_2(z)$ using  standard transformations of the
hypergeometric functions in the following symmetric form 
\begin{equation}\label{f1g}
f_1(z) = \frac{\pi^{3/2}}{\sin 2\pi \Lambda} (-a_1 g_1(z)+b_1 g_2(z))
\end{equation}
\begin{equation}\label{f2g}
f_2(z) =-i\frac{\pi^{3/2}}{\sin 2\pi \Lambda} (-a_2 g_1(z)+b_2 g_2(z))
\end{equation}
where
\begin{equation}\label{g1}
g_1(z) = (1-v)^\Lambda ~_2F_1(\frac{1-\lambda}{4}+\Lambda,
\frac{1+\lambda}{4}+\Lambda,2\Lambda+1,1-v)
\end{equation}
\begin{equation}\label{g2}
g_2(z) = (1-v)^{-\Lambda}~_2F_1(\frac{1-\lambda}{4}-\Lambda,
\frac{1+\lambda}{4}-\Lambda,-2\Lambda+1,1-v)
\end{equation}
and
\begin{equation}\label{a1b1}
a_1= \frac{1}
{\Gamma(1+2\Lambda)\Gamma(\frac{1-\lambda}{4}-\Lambda)
\Gamma(\frac{1+\lambda}{4}-\Lambda)}
,~~~~
b_1= \frac{1}
{\Gamma(1-2\Lambda)\Gamma(\frac{1-\lambda}{4}+\Lambda)
\Gamma(\frac{1+\lambda}{4}+\Lambda)}
\end{equation}
\begin{equation}\label{a2b2}
a_2= \frac{1}
{\Gamma(1+2\Lambda)\Gamma(\frac{3-\lambda}{4}-\Lambda)
\Gamma(\frac{3+\lambda}{4}-\Lambda)},~~~~
b_2= \frac{1}
{\Gamma(1-2\Lambda)\Gamma(\frac{3-\lambda}{4}+\Lambda)
\Gamma(\frac{3+\lambda}{4}+\Lambda)}~.
\end{equation}

Under the cycle $C_2$, using eqs.(\ref{f1g},\ref{f2g}), we have for 
$\Lambda = i B$ the monodromy matrix
for $(f_1,if_2)$
\begin{equation}\label{M(C2)B}
\frac{1}{-a_1b_2+ a_2 b_1}
\begin{pmatrix}
-a_1b_2 e^{2\pi B}+a_2b_1 e^{-2\pi B}
&a_1b_1 (e^{2\pi B}-e^{-2\pi B})\\
a_2b_2 (-e^{2\pi B}+e^{-2\pi B})&a_2b_1 e^{2\pi B}-a_1b_2 e^{-2\pi B}
\end{pmatrix}
\end{equation}
with $b_1 = \bar a_1$, $b_2 = \bar a_2$ and thus
\begin{equation}\label{kappaC2}
\kappa^4 = \frac{a_1b_1}{a_2b_2}=\frac{a_1\bar a_1}{a_2 \bar a_2}>0
\end{equation}
which agrees with the value found from $C_1$. Being 
$-a_1b_2+ a_2 b_1$ pure imaginary we have that the rescaled matrix
belongs to $SU(1,1)$.

For $\Lambda$ real we have for $M(C_2)$
\begin{equation}
\frac{1}{-a_1b_2+ a_2 b_1}
\begin{pmatrix}
-a_1b_2 e^{-2i\pi\Lambda}+a_2b_1 e^{2i\pi\Lambda}
&a_1b_1 (e^{-2i\pi\Lambda}-e^{2i\pi\Lambda})\\
a_2b_2 (-e^{-2i\pi\Lambda}+e^{2i\pi\Lambda})&a_2b_1 e^{-2i\pi\Lambda}-a_1b_2 e^{2i\pi\Lambda}
\end{pmatrix}
\end{equation}
\begin{equation}
\kappa^4 = \frac{a_1b_1}{a_2b_2}
\end{equation}
agreeing with (\ref{kappa4}) and being now $-a_1b_2+ a_2 b_1$ real 
we have again $M_{22}(C_2)=\overline{M_{11}(C_2)}$ and single
valuedness of $\phi$.

In conclusion for any $\Lambda$ imaginary we always have single
valuedness of the conformal factor along the cycles $C_1$ and $C_2$
and the same holds for real but not too large $\Lambda$. We see from
the above formulas that while the monodromy along $C_1$ is always
elliptic with trace $-2 \cos(\pi\lambda)$, 
the monodromy along $C_2$ for $\Lambda=i B$ imaginary is hyperbolic
with trace $2 \cosh(2\pi B)$, while the same monodromy for $\Lambda$
real is elliptic with trace $2 \cos(2\pi\Lambda)$.
 
We can also compute the finite part $X_z$ of the conformal factor at
the origin for $B\neq 0$ to get
\begin{equation}
-X_z(\eta,B) = -\log 6 -2\eta \log\frac{4}{3}+
\log\gamma(\eta+2 i B)+\log\gamma(\eta-2 i B)+2\log\gamma(\frac{1}{2}-\eta)
\end{equation}
which integrated in $\eta$ at fixed $B$, with boundary condition
$G(0,B)=0$ 
provides the following function
which will play a key role in the sequel
\begin{eqnarray}\label{Gfunction}
& &-G(\eta,B)=
-\eta\log 6 -\eta^2\log\frac{4}{3}\\
&+&
F(\eta+2iB)+F(\eta-2iB)-F(2iB)-F(-2iB)-2F(\frac{1}{2}-\eta)~.
\nonumber
\end{eqnarray}

The condition determining the value of the accessory parameter can be
obtained in two different ways. We saw that the reality of the
parameter $\kappa^4$ derived from the monodromy $M(C_1)$ requires $\Lambda$
either real or pure imaginary and from the results after
eq.(\ref{kappa4}) the ensuing conformal factor obeys
$\phi(z)=\phi(\bar z)=\phi(-z)$. 

We can satisfy torus boundary conditions by computing the
monodromy of the transformation $\tilde f_j(z)\equiv
f_j(z-2\omega_2)$, with $\omega_2$ pure imaginary, i.e. the $C_3$ cycle 
(the long cycle) using \cite{torusIII}
\begin{equation}
M_{12}(C_3) = - \tilde f_1(z) f'_1(z)+\tilde f'_1(z) f_1(z),~~~~
M_{21}(C_3) =  \tilde f_2(z) f'_2(z)-\tilde f'_2(z) f_2(z)
\end{equation} 
and imposing
\begin{equation}
\frac{M_{12}(C_3)}{\overline{M_{21}(C_3)}}=\kappa^4
\end{equation}
being $\kappa^4$ given by eq.(\ref{kappa4}).
An alternative amounts to imposing that for $z=x+iy$ at
$y = -i\omega_2$ the derivative of 
$e^{-\phi/2}$ w.r.t. $y$ vanishes i.e. given
\begin{equation}
e^{-\phi/2} = {\rm const}~ [\kappa^{-2}\bar f_1(\bar z) f_1(z)-
\kappa^{2}\bar f_2(\bar z) f_2(z)]
\end{equation}
\begin{equation}\label{vanishingderivative}
0=
\kappa^{-2}\big(\bar f'_1(\bar z) f_1(z)-f_1(\bar z) f'_1(z)\big)-
\kappa^{2}\big(\bar f'_2(\bar z) f_2(z)-f_2(\bar z) f'_2(z)\big)
\end{equation}
thus allowing for a solution $\phi(z)$ periodic in $y$.
The two methods, as expected give the same result. In fact the
functions $f_1,f_2$ are real analytic functions of $z$. This is
apparent for $\Lambda$ real. For $\Lambda$ pure imaginary if follows
from the fact that they satisfy eq.(\ref{zdiffeq}) with $q^2=0$ 
with a real $\beta$ and real boundary conditions at $z=\omega_1$, 
$f_1(\omega_1)$=1,~$f_1'(\omega_1)$=0,~ 
$f_2(\omega_1)$=0,~$f'_2(\omega_1)=\pi/\omega_1$. Then we have at
$y=-i\omega_2$ 
\begin{equation}
f_1(z)f_1'(z)|_{z=\bar z}-f'_1(z)f_1(z)|_{z=\bar z}=
f_1(z)\bar f_1'(\bar z)-f'_1(z)\bar f_1(\bar z)~.
\end{equation}
It is useful to introduce the variable
$\zeta$ given by $z= \omega_1(1+\zeta)$, $\zeta = t_1+it$ and write 
$T= e^{i\pi\zeta}$;
for large and positive $t={\rm Im}~\zeta$, $T$ tends to
zero. Recalling that $1-v = [\cos(\pi\zeta/2)]^{-2}$,
for $g_1$ and
$g_2$ we have the following expansion convergent in ${\rm Im}\zeta>0$
\begin{equation}\label{g1Texpansion}
g_1 = 4^\Lambda T^\Lambda \bigg(1+\epsilon\frac{1}{1+2\Lambda}~T
-\epsilon ~\frac{1}{2~(1+\Lambda)}~T^2+
\epsilon^2\frac{1}{4(1+\Lambda)(1+2\Lambda)}~T^2 +O(T^3)\bigg)
\end{equation}
\begin{equation}\label{g2Texpansion}
g_2 = 4^{-\Lambda} T^{-\Lambda} \bigg(1+\epsilon\frac{1}{1-2\Lambda}~T
-\epsilon ~\frac{1}{2~(1-\Lambda)}~T^2+
\epsilon^2\frac{1}{4(1-\Lambda)(1-2\Lambda)}~T^2 +O(T^3)\bigg)
\end{equation}
where $g_2$ is simply obtained from $g_1$ by sending $\Lambda$ into $-\Lambda$.

The above given expressions are very useful to compute the large 
${\rm Im}\zeta$ behavior of the $f_k$. As in this section we work 
$O(q^0)$ only the first term in the expansion intervenes. For
$\Lambda$ real we obtain from eq.(\ref{vanishingderivative})
\begin{equation}
\kappa^{-2} (a_1^2 4^{2\Lambda}e^{i\pi\Lambda(\zeta-\bar\zeta)}-
b_1^2 4^{-2\Lambda}e^{-i\pi\Lambda(\zeta-\bar\zeta)})=
\kappa^{2} (a_2^2 4^{2\Lambda}e^{i\pi\Lambda(\zeta-\bar\zeta)}-
b_2^2 4^{-2\Lambda}e^{-i\pi\Lambda(\zeta-\bar\zeta)})
\end{equation}
which using eq.(\ref{kappa4}) gives
\begin{equation}
4^{4\Lambda}e^{-4\pi \Lambda t}= -\frac{b_1b_2}{a_1a_2}~.
\end{equation}
As we are interested in the limit of large $t$, such an equation
should be solvable for small $\Lambda$ but this is
not possible as the left hand side is always positive while the right
hand side for $\Lambda \rightarrow 0$ tends to $-1$. We conclude that
an elliptic deformation even if it can satisfy the monodromy condition
along the cycles $C_1$ and $C_2$ it cannot satisfy the monodromy
condition along the cycle $C_3$.

We examine now the case of imaginary $\Lambda$, $\Lambda \equiv i
B$, i.e. the hyperbolic deformation. In this case we find
\begin{equation}
\kappa^{-2} (a_1^2 4^{2 i B}e^{-\pi B (\zeta-\bar\zeta)}-
b_1^2 4^{-2i B}e^{\pi B(\zeta-\bar\zeta)})=
\kappa^{2} (a_2^2 4^{2 i B}e^{-\pi B (\zeta-\bar\zeta)}-
b_2^2 4^{-2i B}e^{\pi B(\zeta-\bar\zeta)})
\end{equation}
which using eq.(\ref{kappaC2}) gives
\begin{equation}\label{Bequation}
4^{4 i B}e^{-4i\pi B t} = -\frac{\bar a_1\bar a_2}{a_1a_2}~.
\end{equation}
From the expressions (\ref{a1b1},\ref{a2b2}) for the $a_j, b_j$ and Legendre
duplication formula
$\Gamma(2z)=2^{2z-1}\Gamma(z)\Gamma(z+1/2)/\sqrt{\pi}$
the r.h.s. of equation (\ref{Bequation}) can be written as
\begin{equation}\label{duplication}
-\frac{\Gamma^2(1+2iB)\gamma(\frac{1-\lambda}{2}-2iB)}
{\Gamma^2(1-2iB)\gamma(\frac{1-\lambda}{2}+2iB)}
e^{4iB\log 4}
\end{equation}
thus obtaining the equation for $B$ as a function of $\omega_2 =
i t \omega_1=i\tau_2\omega_1$
\begin{equation}\label{lowestorder}
4\pi Bt = i\log\bigg[-\frac{\Gamma^2(1+2iB)\gamma(\frac{1-\lambda}{2}-2iB)}
{\Gamma^2(1-2iB)\gamma(\frac{1-\lambda}{2}+2iB)}\bigg]=
\pi+ i\log\frac{\Gamma^2(1+2iB)\gamma(\eta-2iB)}
{\Gamma^2(1-2iB)\gamma(\eta+2iB)}~.
\end{equation}
The $\pi$ appearing on the r.h.s. of eq.(\ref{lowestorder}),
originates from the minus sign in eq.(\ref{duplication}), $i\log(-1)= \pi$. 
We could have also
chosen $-\pi$; then the solution of eq.(\ref{lowestorder}) will simply
be minus the solution of the previous equation. This is due to the
fact that the solution of the problem i.e. the conformal factor, is
invariant under $B\rightarrow -B$. Higher values of $i\log(-1)$ like 
$3\pi$ correspond to solutions of eq.(\ref{vanishingderivative}) lying 
beyond the first hyperbolic horizon. From the value of $B$
we can also compute the trace of the $C_3$ monodromy. For
large $t$ we find 
\begin{equation}
M_{11}(C_3)=M_{22}(C_3) = \frac{1}{2\pi B} \cos\frac{\pi\lambda}{2}
\end{equation}
where according to eq.(\ref{lowestorder}) $B = 1/4t+O(1/t^2)$ 
and as such it is an hyperbolic monodromy.

Given the value of $B$ extracted from eq.(\ref{lowestorder}) 
one can compute the
action to order $O(q^0)$ i.e. keeping into account all
logarithmic corrections  by using the relation
\begin{equation}\label{torusaction}
\frac{1}{2\pi} S_z(\eta)= -\int_0^\eta X_z(\eta, B(\eta))d\eta
\end{equation}
where $X_z$ is the finite part of the conformal factor $\phi$ at the
origin
\begin{equation}
\phi = -2\eta \log|z|^2 + X_z + o(z)~.
\end{equation}
The action (\ref{torusaction}) is really the action on the torus
because if one
limits the integration of $\frac{1}{2}\partial_z\phi\partial_{\bar z}\phi+
e^\phi$ to the periodicity region $-\omega_1<{\rm Re}~z<\omega_1$,  
$-\tau_2\omega_1<{\rm Im}~z<\tau_2\omega_1$ 
we have no boundary terms in the action and the only source of variation
of $S_z$ is just the contribution at the origin $X_z$. The derivative
of the function $G$ eq.(\ref{Gfunction}) with respect to $B$ gives 
\begin{equation}
\frac{\partial G}{\partial B}= 2 i \log\bigg(-\frac
{\Gamma(1+2iB)^2\gamma(\eta-2iB)}{\Gamma(1-2iB)^2\gamma(\eta+2iB)}\bigg)
\end{equation}
which is twice the r.h.s. of eq.(\ref{lowestorder}). 
Then we see that the $S_z/2\pi$
of eq.(\ref{torusaction}) coincides with the value of the expression
\begin{equation}\label{minimum}
4\pi B^2 t -G(\eta,B)
\end{equation}
computed at the value of $B(\eta)$ which realizes the minimum of
eq.(\ref{minimum}).
In fact we have for the total derivative of eq.(\ref{minimum}) with respect 
to $\eta$ 
\begin{equation}
\bigg(8\pi B(\eta)~ t -\frac{\partial G(\eta,B(\eta)}{\partial B}\bigg)
\frac{d B}{d \eta}-
\frac{\partial G(\eta,B(\eta))}{\partial\eta}=
-X_z(\eta,B(\eta))~.
\end{equation}
We give below the comparison with the semiclassical limit of the quantum
one-point function.
The primary fields in Liouville theory are given by
$e^{2\alpha\phi(z,\bar z)}$. 
Following the notation of
\cite{hadaszrecursive}, but replacing $\lambda$ with $l$ not to create
confusion with the $\lambda$ introduced in the previous sections of the
present paper, we have for the
dimension $\Delta$ of $V_{l,l} = e^{2\alpha \phi(z,\bar z)}$
\begin{equation}
\Delta = \alpha(Q-\alpha)=\frac{1}{4}(Q^2-l^2)~~~~{\rm where}~~~~
\alpha = \frac{Q}{2}+
\frac{l}{2}~.
\end{equation}
The central charge in Liouville theory is given by
\begin{equation}
c = 1 + 6 Q^2~,~~~~~~~~ Q = \frac{1}{b}+b ~.
\end{equation}
The torus one-point function is given by \cite{hadaszrecursive,albamorozov,DMS}
\begin{equation}\label{Trace}
\langle V_{l,\bar l}\rangle = 
{\rm Tr} (e^{-\tau_I \hat H+ i \tau_R \hat P}
V_{l,\bar l}(1,1))
=(\tilde q\bar{\tilde q})^{-\frac{c}{24}}
{\rm Tr} (\tilde q^{L_0}\bar{\tilde q}^{\bar L_0} V_{l,\bar l}(1,1))
\end{equation}
where
${\tilde q}=e^{2\pi i\tau}$.
The trace has to be computed on the Verma module
\begin{equation}
\nu_{\Delta,M} = L_{-M}\nu_\Delta=L_{-m_j}\dots
L_{-m_1}\nu_{\Delta},~~~~
\bar\nu_{\Delta,\bar N} = \bar L_{-\bar N}\bar\nu_\Delta=\bar L_{-n_j}\dots
\bar L_{-n_1}\bar\nu_{\Delta}
\end{equation}
with \cite{DMS} 
\begin{equation}\label{matrixelement}
\langle V_1|V_3(z,\bar z)|V_2\rangle=\lim_{w\rightarrow \infty}
w^{2\Delta_1}\bar w^{2\bar\Delta_1}\langle V_1(w,\bar w)V_3(z,\bar z) 
V_2(0,0)\rangle ~.
\end{equation}
Only matrix elements with the same dimensions 
appear in the computation of the conformal
block. In Liouville we have real $\Delta$ and $\Delta=\bar\Delta$. 
The fundamental matrix element (\ref{matrixelement})
\begin{equation}
\langle \nu_{\Delta'},\bar\nu_{\Delta'},|V_{l,\bar l}(1,1)|
\nu_{\Delta'},\bar\nu_{\Delta'}\rangle = C(\frac{Q}{2}-\frac{l'}{2},
\frac{Q}{2}+\frac{l}{2},\frac{Q}{2}+\frac{l'}{2})
\end{equation}
is provided by the DOZZ \cite{dornotto,ZZ,teschner} structure constant.
Due to the continuum spectrum of Liouville theory the general formal
expression \cite{hadaszrecursive} for the trace (\ref{Trace}) 
($[B^n_{c,\Delta}]^{MN}$ are the inverses of the Kac matrices)
\begin{eqnarray}
\langle V_{l,\bar l}\rangle&=&
(\tilde q\bar{\tilde q})^{-\frac{c}{24}}
\sum_{\Delta,\bar\Delta}\sum_{n=0}^\infty
\sum_{m=0}^\infty {\tilde q}^{\Delta+n} \bar{\tilde q}^{\bar\Delta+m}
\sum_{n=|M|=|N|,m=|\bar M|=|\bar N|} [B^n_{c,\Delta}]^{MN}
[\bar B^m_{c,\bar\Delta}]^{\bar M \bar N} \nonumber\\ 
&&\times\langle\nu_{\Delta,M}\nu_{\bar\Delta,\bar M}
|V_{l,\bar l}(1,1)|\nu_{\Delta,N}\nu_{\bar\Delta,\bar N}\rangle
\end{eqnarray}
goes over to 
\begin{equation}\label{torusonepoint}
\langle V_{l,\bar l}\rangle = \int_0^{i\infty} \frac{d l'}{i}
{\cal F}^{l}_{c,\Delta'}(\tilde q){\cal F}^{l}_
{c,\bar\Delta'}(\bar {\tilde q})
C^{l,l}_{\Delta',\Delta'}=
\int_0^{i\infty} \frac{d l'}{i}
|{\cal F}^{l}_{c,\Delta'}(\tilde q)|^2
C(\frac{Q}{2}-l',\frac{Q}{2}+\frac{l}{2},\frac{Q}{2}+\frac{l'}{2})
\end{equation}
with 
\begin{equation}\label{torusonepoint2}
{\cal F}^{l}_{c,\Delta'}(\tilde q) = 
{\tilde q}^{\Delta'-\frac{c}{24}}\sum_{n=0}^\infty
{\tilde q}^n F^{l,n}_{c,\Delta'}
\end{equation}
the conformal block. Such $1$-point torus conformal blocks are not yet
known in general form \cite{albamorozov,mironov}.

The semiclassical limit is obtained for $b\rightarrow 0$.
Using $l' = \tilde l_p/b$, $\tilde l_p =ip$ and $l=\tilde l/b$ we have
\begin{equation}
\Delta' = \frac{1}{4}(Q^2-l'^2)\rightarrow \frac{1}{b^2}\frac{1}{4}
(1-\tilde l_p^2)=\frac{1}{b^2}\frac{1}{4}(1+p^2)
\end{equation}
and \cite{ZZ}
\begin{equation}
C(\frac{Q}{2}-l',\frac{Q}{2}+\frac{l}{2},\frac{Q}{2}+\frac{l'}{2})
\rightarrow e^{-\frac{S^{(cl)}}{2\pi b^2}}
\end{equation}
\begin{equation}\label{Scl}
\frac{S^{(cl)}}{2\pi}= F(\eta)+F(\eta+ip)+F(\eta-ip)+F(1-\eta)
-F(1+ip)-F(2\eta)-F(1-ip)-\eta \log 2
\end{equation}
where $\eta=(1+\tilde l)/2 = (1-\lambda)/2$.
The equation for the saddle point to order $q^0$ is
\begin{equation}
-4\pi\tau_2~\frac{1}{2}p -\frac{1}{2\pi}\frac{\partial S^{(cl)}}{\partial p}=0
\end{equation}
where recalling the definition of the function $F(\eta)$ eq.(\ref{F}) 
\begin{equation}\label{dScldp}
\frac{1}{2\pi}\frac{\partial S^{(cl)}}{\partial p}
= i\big(\log \gamma(\eta+ip)-\log \gamma(\eta-ip)-
\log\gamma(1+ip)+\log\gamma(1-ip)\big)
\end{equation}
and the equation for $p$ becomes
\begin{equation}\label{pequation}
2\pi \tau_2 p = i\log\bigg(-\frac{\Gamma^2(1+ip)\gamma(\eta-ip)}
{\Gamma^2(1-ip)\gamma(\eta+ip)}\bigg)
\end{equation}
which is the same as eq.(\ref{lowestorder}). One
sees from eqs.(\ref{lowestorder}) and (\ref{pequation}) that the 
correspondence between the classical accessory parameter $B$ and the
saddle point value $p_s$ of $p$
is $p_s=2 B$. With regard to the value of the action computed from the saddle
point we have from eq.(\ref{Scl})
\begin{eqnarray}\label{dScldeta}
\frac{1}{2\pi}\frac{\partial S^{(cl)}}{\partial \eta}&=& 
\log\gamma(\eta)-\log\gamma(1-\eta)-2\log\gamma(2\eta)+
\log\gamma(\eta+ip)+\log\gamma(\eta-ip)-\log 2\nonumber\\
&=&-X_z-(2\eta-1)\log 12 ~.
\end{eqnarray}
The term $(2\eta-1)\log 12$ is due to our choice
$\omega_1=\pi/(2\sqrt{3})$, see section \ref{general},
instead of the standard one $\omega_1=\pi$, and the
scaling behavior of the 1-point function whose classical dimension is
given by $\eta(1-\eta)$. 
Thus also the action computed from the saddle
point agrees with action computed from the imposition of the monodromy
along the cycle $C_3$.

\section{Discussion of higher order calculations}\label{discussion}
As we mentioned already in the introduction even if the terms
$K_n$ appearing in the expansion of $K$ eq.(\ref{Kexpansion}) are formally 
of order $O(q^{2n})$
due to the exponential behavior of $\cos(n\pi z/\omega_1)$ at large
imaginary values of $z$, such term contributes in the corrections to 
the $f_j$ to order $O(q^n)$. In particular for $n=1$ we shall have $O(q)$
contributions.
This appears in contrast with the structure of the quantum one-point
function (\ref{torusonepoint},\ref{torusonepoint2}) 
where only $q^2\equiv \tilde q$ appears. 
We show in the following that such $O(q)$ contributions cancel exactly
in the equation determining the value of the accessory parameter.
The corrections $O(q^2)$ have several origins. In imposing the
$SU(1,1)$ nature of the monodromy along the cycle $C_3$ 
or equivalently in imposing  eq.(\ref{vanishingderivative}), 
the functions $f_k(z)$ at ${\rm Im}z=\omega_1 t$ will have to
be computed to order $T$ and $T^2$ included expanding the
hypergeometric function
appearing in (\ref{g1},\ref{g2}) according to eqs.(\ref{g1Texpansion},
\ref{g2Texpansion}). 
Secondly to reach order $O(q^2)$ two 
iterations in the solution of the integral equation (\ref{volterra}) 
have to be performed.

First we must examine the change in the parameter $\kappa^4$ due to
the kernel appearing in eq.(\ref{volterra}). At the
origin $z\approx 0$ the corrected solutions $f^{(1)}_k$ behave, 
in matrix form, as
\begin{equation}
f^{(1)} = (1+\Delta)f
\end{equation}
where
\begin{equation}
\Delta =-2\pi\epsilon q^2
\begin{pmatrix}
I_{12}&-I_{11}\\
I_{22}&-I_{12}
\end{pmatrix}
\end{equation}
being $I_{jk}$ the numbers
\begin{equation}
I_{jk} = \int_0^{-1} \cos\pi\zeta ~f_j(\zeta)~f_k(\zeta)~d\zeta~.
\end{equation}
The monodromy matrix $M(C_1)$ goes over to
\begin{equation}\label{M^1C1}
M^{(1)}(C_1)=M(C_1)+[\Delta,M(C_1)]~.
\end{equation}
The $I_{jk}$ are all real and this assures through the nature of the
$M(C_1)$ matrix that $M^{(1)}_{22}(C_1)=\overline{M^{(1)}_{11}(C_1)}$
and that the ratio $M^{(1)}_{12}(C_1)/\overline{M^{(1)}_{21}(C_1)}$ is
real and positive. Such a ratio provides the value of the corrected $\kappa^4$
\begin{equation}\label{kappa4I}
\kappa^4\bigg(1-2\pi\epsilon q^2\big[4
I_{12}+(M_{11}(C_1)-M_{22}(C_1))(\frac{I_{11}}{M_{12}(C_1)}+\frac{I_{22}}
{\overline{M_{21}(C_1)}})\big]\bigg)~.
\end{equation}
In addition it is easy to prove from eq.(\ref{M^1C1}) applied
to the half-turn monodromy matrix, that the new corrected conformal
factor still satisfies $\phi(-z)=\phi(z)$.

The monodromy is still elliptic with trace $-2 \cos(\pi \lambda)$.
There is an alternative way to compute such a correction, i.e. by 
imposing the $SU(1,1)$ nature of the monodromy along the cycle $C_2$.
This is important as it enlightens the structure of the functions
$g_k$ and their $O(q^2)$ corrections for large values of ${\rm
Im}\zeta$ which is the region of interest in computing the monodromy
along the cycle $C_3$, i.e. the long cycle and sets a relation between
the $I_{jk}$ and other integration constants.

To this end we start from the two functions 
\begin{equation}
f_1(z) = \frac{\pi^{3/2}}{\sin 2\pi \Lambda} (-a_1 g_1(z)+b_1 g_2(z))
\end{equation}
\begin{equation}
if_2(z) =\frac{\pi^{3/2}}{\sin 2\pi \Lambda} (-a_2 g_1(z)+b_2 g_2(z))
\end{equation}
which are canonical at $z=\omega_1$ and compute the corrections on the
$g_k$. We have this time, in matrix form,
\begin{equation}\label{gcorr}
g^{(1)}(\zeta) = (1+R(\zeta))g(\zeta)
\end{equation}
where taking into account the wronskian factor
\begin{equation}
R(\zeta) =\frac{\pi\epsilon q^2}{B}
\begin{pmatrix}
J_{12}(\zeta)&-J_{11}(\zeta)\\
J_{22}(\zeta)&-J_{12}(\zeta)
\end{pmatrix}
\end{equation}
with
\begin{equation}\label{Jintegrals}
J_{jk}(\zeta) = \int_0^{\zeta} \cos\pi\zeta' ~g_j(\zeta')
~g_k(\zeta')~d\zeta'~.
\end{equation}
The structure of the $J_{kl}(\zeta)$ is
\begin{equation}
J_{12}(\zeta) = s_{12}(e^{i\pi\zeta})+\frac{\epsilon\zeta}{1+4B^2}+r_{12}
\end{equation}
\begin{equation}
J_{11}(\zeta) = 4^{2iB}e^{-2\pi B\zeta}~s_{11}(e^{i\pi\zeta})+r_{11}
\end{equation}
\begin{equation}
J_{22}(\zeta) = 4^{-2iB}e^{2\pi B\zeta}~s_{22}(e^{i\pi\zeta})+r_{22}
\end{equation}
where the $s_{jk}$ being functions of $e^{i\pi\zeta}$ are invariant
under $\zeta\rightarrow \zeta-2$ and the numbers $r_{jk}$
are the contributions due to the lower integration limit
in eq.(\ref{Jintegrals}). From
eq.(\ref{g1Texpansion},\ref{g2Texpansion}) we have
$r_{12}$ pure imaginary and $r_{22} = -\bar
r_{11}$. Thus we can write to order $q^2$ the corrected $g_1$ as
\begin{eqnarray}
&&g_1(\zeta)~e^{\frac{\epsilon^2
    q^2\pi\zeta}{B(1+4B^2)}}\bigg(1+\frac{\epsilon\pi q^2}{B}\big[
s_{12}(e^{i\pi\zeta})+r_{12}-
4^{2iB} e^{-2\pi B\zeta}s_{11}(e^{i\pi\zeta})g_2(\zeta)\big]\bigg)-
\frac{\epsilon\pi q^2}{B}r_{11}~g_2(\zeta) \equiv\nonumber\\
&&\tilde g_1(\zeta)-\frac{\epsilon\pi q^2}{B}r_{11}~g_2(\zeta)
\end{eqnarray}
and similarly for $g_2$. Under the cycle $C_2$ i.e. $\zeta\rightarrow
\zeta-2$ (the $C_2$ monodromy can be computed at any value
of ~${\rm Im}\zeta$) the $\tilde g_j$ transform like $\tilde g_1\rightarrow 
e^{2\pi B'}\tilde g_1$ and $\tilde g_2\rightarrow e^{-2\pi B'}\tilde g_2$ with
\begin{equation}
B'= B-\frac{\epsilon^2 q^2}{B(1+4B^2)}~.
\end{equation}
Denoting by ${\cal B}'$, $R$ and $A$ the matrices
\begin{equation}\label{Bprime}
{\cal B}'= 
\begin{pmatrix}
e^{2\pi B'}&0\\
0&e^{-2\pi B'}
\end{pmatrix},~~~~
R= \frac{\epsilon\pi q^2}{B}
\begin{pmatrix}
0 &-r_{11}\\
r_{22} & 0
\end{pmatrix},
~~~~~A= 
\begin{pmatrix}
-a_1&b_1\\
-a_2&b_2
\end{pmatrix}
\end{equation}
we have for the $C_2$ monodromy to order $q^2$
\begin{equation}\label{M(C3)J}
M^{(1)}(C_2) = A{\cal B}'A^{-1}+A[R,{\cal B}]A^{-1}
\end{equation}
where ${\cal B}$  is the matrix ${\cal B}'$ with $B'$ replaced by
$B$.
Due to the diagonal nature of ${\cal B}'$, the ratio of the 
off-diagonal elements of $A{\cal B}'A^{-1}$
is the same as those of the unperturbed monodromy
eq.(\ref{M(C2)B}). For the second term we find
\begin{equation}
2\frac{\epsilon\pi q^2 \sinh(2 \pi B)}{B \det A }
\begin{pmatrix}
-a_1a_2r_{11}-\bar a_1\bar a_2\bar r_{11}&a_1^2 r_{11}+\bar a_1^2
\bar r_{11}\\
-a_2^2 r_{11}-\bar a_2^2\bar r_{11}&a_1a_2r_{11}+\bar a_1\bar a_2\bar r_{11}
\end{pmatrix}
\end{equation} 
where $\det A=-a_1\bar a_2+\bar a_1 a_2$. Again taking the ratio 
$M^{(1)}_{12}(C_2)/\overline{M^{(1)}_{21}(C_2)}$ we have the corrected
value of $\kappa^4$
\begin{equation}\label{kappa4J}
\kappa^4\bigg(1+2\frac{\pi\epsilon q^2 \sinh(2 \pi B)}{B \det A}\big[
\frac{a_1^2 r_{11}+\bar a_1^2 \bar r_{11}}{M_{12}(C_2)}-
\frac{a_2^2 r_{11}+\bar a_2^2 \bar r_{11}}{\overline{M_{21}(C_2)}}\big]\bigg)~.
\end{equation}
Moreover from eq.(\ref{M(C3)J}) we read for the trace of the $C_2$
monodromy the value ${\rm Tr}M^{(1)}(C_2) = 2 \cosh(\pi B')$, correct
to order $O(q^2)$ included.

The two values of $\kappa^4$ eq.(\ref{kappa4I}) and eq.(\ref{kappa4J}) 
have to agree due to the fact that the 
invariance of the conformal factor under inversion $z\rightarrow -z$
and under reflections ${\rm Im}z\rightarrow-{\rm Im}z$ imply monodromy 
under $C_2$. This sets a relation between the value
of $r_{11}$ arising from the lower integration limit in eq.(\ref{Jintegrals})
and the integrals $I_{jk}$ of product of hypergeometric functions. We
verified the validity of such a relation with high precision numerical
tests.

We come now to the resolution of the apparent contradiction of the
presence of $O(q)$ corrections for the equation determining the
accessory parameter $B$.

In deriving the equation for the accessory parameter one
needs to compute the functions $f_k$ i.e. the $g_k$ for 
${\rm Im}\zeta=\tau_2$ i.e. for large imaginary values of
$\zeta$.
The explicit form of such corrections due to the first iteration 
with $K_1$ in
equation (\ref{volterra}) is given using eq.(\ref{gcorr}) by  
\begin{eqnarray}\label{deltag1}
& & \delta g_1 = K_1g_1\\
&=&\epsilon q^24^{i B} e^{-\pi B\zeta}\bigg(
\frac{e^{-i\pi \zeta}}{1-2i B}+
\epsilon\frac{\pi \zeta }{B(1+4B^2)}
+{\rm const}\bigg) -\frac{\pi\epsilon q^2}{B}~4^{-iB}r_{11}e^{\pi B\zeta}+
O(e^{-\pi{\rm Im}(\zeta)})\nonumber
\end{eqnarray}
and similarly
\begin{eqnarray}\label{deltag2}
& & \delta g_2 = K_1g_2\\
&=&\epsilon q^24^{-i B} e^{\pi B\zeta}\bigg(
\frac{e^{-i\pi \zeta}}{1+2i B}-
\epsilon\frac{\pi \zeta }{B(1+4B^2)}
+{\rm const}\bigg) +\frac{\pi\epsilon q^2}{B}~4^{iB}r_{22}e^{-\pi B\zeta}+
O(e^{-\pi{\rm Im}(\zeta)})~.\nonumber
\end{eqnarray}
In the computation of the $C_3$ monodromy after taking the
derivatives in eq.(\ref{vanishingderivative}) one has to set 
${\rm Im}\zeta=\tau_2$.
Thus we see from the presence of the factor $e^{-i\pi \zeta}$ that the
first terms in eq.(\ref{deltag1},\ref{deltag2}) contribute $q^2 q^{-1}=q$ and
this seems to contrast with the structure of the quantum one-point
function where only $\tilde q \equiv q^2$ appears. However the second
term in the expansion of $g_1$ in (\ref{g1Texpansion})
\begin{equation}
\epsilon~4^{iB} e^{-\pi B \zeta} \frac{e^{i\pi\zeta}}{1+2iB}
\end{equation}
is also of order $O(q)$ and the two contributions cancel exactly 
in the computation of the $C_3$ monodromy thus leaving only
$q^2$ terms in the problem. However the complete evaluation of the $q^2$
corrections to the monodromy implies the computation of the iterated 
$K_1\times K_1$ contribution and of $K_2$ which is a rather lengthy
process.
\section{Conclusions}
In the present paper we developed a method for computing the accessory
parameter for the torus with one generic source in the regime of high
values of the imaginary part of the modulus. This is the region which
is of interest in the usual formulation of the conformal blocks in
quantum conformal theories. The main idea is to use an
expansion of Weierstrass $\wp$ function in the parameter
$q=e^{i\pi\tau}$. The advantage in that the zero order problem is
related to a soluble one i.e. the infinite strip, and actually turns
out to be an hyperbolic deformation of it, which is also soluble.
In this way we reproduce to $O(q^0)$  
the same equation for the accessory parameter and the same value for
the action as those obtained from
the saddle point method applied to the semiclassical limit of the
quantum one-point function. In principle the procedure can be carried
over to all orders in $q$. We also have given
the full $O(q^2)$ contribution to the change of the weight parameter
$\kappa^4$ necessary to extract the equation for the
accessory parameter to $O(q^2)$, and performed the first iteration of
the Volterra equation. As discussed in the text the procedure
generates also terms $O(q)$
which are absent in the expression of the quantum one-point function
but it is shown that they give contributions which 
cancel exactly. Due to the
behavior of the kernel of the Volterra equation for large imaginary
values of the coordinate, to reach the order $O(q^{2n})$ one needs not
$n$ but $2n$ iterations of the integral equation.
Thus a complete computation of the $O(q^2)$ terms is already 
a lengthy procedure as it involves the second iteration of
the Volterra with the kernel $K_1$ and one iteration with the kernel
$K_2$ as both give $O(q^2)$ contributions to the equation for $B$ 
and such computation will be attempted elsewhere. 
\section*{Acknowledgments}
The author is grateful to Massimo Porrati for correspondence.

\vfill


\end{document}